\documentclass[conference,10pt]{IEEEtran}
\usepackage{amsmath,amsfonts}
\usepackage{algorithmic} 
\usepackage{array}
\usepackage{textcomp}
\usepackage{stfloats}
\usepackage{url}
\usepackage{verbatim}
\usepackage{graphicx}

\usepackage{balance}

\usepackage{amssymb}
\usepackage{cite}
\usepackage{xcolor}
\usepackage[inline]{enumitem}
\usepackage{enumerate}
\usepackage{subcaption} 
\usepackage{hyperref}
\usepackage{listings}
\usepackage{pifont}
\usepackage{pythonhighlight}
\usepackage[ruled,vlined,linesnumbered]{algorithm2e}
\usepackage{titlesec}
\usepackage{orcidlink}
\usepackage{multirow}

\newcommand{\SynthCount}{$261$~}

\newcommand{\SynthSpeedUp}{$1.9\times$~}

\newcommand{\uptoModelSpeedUp}{$3\times$~}
\newcommand{\uptoModelEnergyreduction}{$2.4\times$~}

\newcommand{\dualModelSpeedUp}{$2.4\times$~}
\newcommand{\dualModelEnergyreduction}{$1.7\times$}

\newcommand{\dualModelTCONVSpeedUp}{$2.7\times$~}

\newcommand{\atleastGOPsDSPx}{$2\times$~}

\definecolor{codegreen}{rgb}{0,0.6,0}
\definecolor{codegray}{rgb}{0.5,0.5,0.5}
\definecolor{codepurple}{rgb}{0.58,0,0.82}
\definecolor{backcolour}{rgb}{0.95,0.95,0.92}

\lstdefinestyle{mystyle}{
	backgroundcolor=\color{white},
	keywordstyle=\color{codegreen},
	numberstyle=\tiny\color{codegray},
	stringstyle=\color{codepurple},
	basicstyle=\ttfamily,
	breakatwhitespace=false,
	breaklines=true,
	captionpos=b,
	keepspaces=true,
	numbers=left,
	numbersep=5pt,
	showspaces=false,
	showstringspaces=false,
	showtabs=false,
	tabsize=2,
    morekeywords=[1]{
    },
}

\lstdefinestyle{customMLIR}{
    inputencoding=utf8,
    tabsize=2,
    rulecolor=,
    upquote=true,
    columns=fixed,
    linewidth=\columnwidth,
    showstringspaces=false,
    extendedchars=true,
    breaklines=true,
    showtabs=false,
    showspaces=false,
    showstringspaces=false,
    basicstyle=\scriptsize\ttfamily,
    identifierstyle=\scriptsize\ttfamily,
    keywordstyle=\scriptsize\ttfamily\color[rgb]{0,0,1},
    commentstyle=\scriptsize\ttfamily\color[rgb]{0.133,0.545,0.133},
    stringstyle=\scriptsize\ttfamily\color[rgb]{0.627,0.126,0.941},
}

\makeatletter

\lstdefinelanguage{mlir}{
  morecomment = [l]{//},
  morestring=[b]', 
  sensitive = true,
  classoffset=0,
  classoffset=1, keywordstyle=\color{purple},
  morekeywords={
    soda.launch_func,
    soda.module,
    soda.return,
    func,
    define, declare, global, constant,
    internal, external, private,
    linkonce, linkonce_odr, weak, weak_odr, appending,
    common, extern_weak,
    thread_local, dllimport, dllexport,
    hidden, protected, default,
    except, deplibs,
    volatile, fastcc, coldcc, cc, ccc,
    x86_stdcallcc, x86_fastcallcc,
    ptx_kernel, ptx_device,
    signext, zeroext, inreg, sret, nounwind, noreturn,
    nocapture, byval, nest, readnone, readonly, noalias, uwtable,
    inlinehint, noinline, alwaysinline, optsize, ssp, sspreq,
    noredzone, noimplicitfloat, naked, alignstack,
    module, asm, align, tail, to,
    addrspace, section, alias, sideeffect, c, gc,
    target, datalayout, triple,
    blockaddress,
    return,
    step,
    func.func, scf.for, 
    linalg.matmul, linalg.generic,
    memref.load, memref.store, memref.subview,
    addi, muli, addf, mulf
  },
  classoffset=2, keywordstyle=\color{gray},
  morekeywords={
    memref, index, f32,
    !iarr_t, !varr_t,
    !AH_t, !AHW_t,
    !mr4x4_0, !mr4x4_1
  },
  alsoletter={\%, ., \!},
  keywordsprefix={\%},
}
\makeatother

\def\BibTeX{{\rm B\kern-.05em{\sc i\kern-.025em b}\kern-.08em
    T\kern-.1667em\lower.7ex\hbox{E}\kern-.125emX}}

\begin{document}
\bstctlcite{IEEEexample:BSTcontrol}

\title{Accelerating Transposed Convolutions on FPGA-based Edge Devices}


\author{Jude Haris, Jos\'e Cano \\
\emph{School of Computing Science, University of Glasgow, Scotland, UK}
}





\maketitle


\begin{abstract}

Transposed Convolutions (TCONV) enable the up-scaling mechanism within generative Artificial Intelligence (AI) models.
However, the predominant Input-Oriented Mapping (IOM) method for implementing TCONV has complex output mapping, overlapping sums, and ineffectual computations.
These inefficiencies further exacerbate the performance bottleneck of TCONV and generative models on resource-constrained edge devices. 
To address this problem, in this paper we propose MM2IM, a hardware-software co-designed accelerator that combines Matrix Multiplication (MatMul) with col2IM to process TCONV layers on resource-constrained edge devices efficiently.
Using the SECDA-TFLite design toolkit, we implement MM2IM and evaluate its performance across \SynthCount TCONV problem configurations, achieving an average speedup of \SynthSpeedUp against a dual-thread ARM Neon optimized CPU baseline. 
We then evaluate the performance of MM2IM on a range of TCONV layers from well-known generative models achieving up to $4.2\times$ speedup, and compare it against similar resource-constrained TCONV accelerators, outperforming them by at least \atleastGOPsDSPx GOPs/DSP.
Finally, we evaluate MM2IM on the DCGAN and pix2pix GAN models, achieving up to \uptoModelSpeedUp speedup and \uptoModelEnergyreduction energy reduction against the CPU baseline.

\end{abstract}


\begin{IEEEkeywords}
    DNN Accelerators, FPGAs, TCONV, Edge AI.
\end{IEEEkeywords}

\section{Introduction}

Generative Artificial Intelligence (AI) is used in various applications, including image-super resolution~\cite{dongAcceleratingSuperResolutionConvolutional2016}, style transfer~\cite{johnsonPerceptualLossesRealTime2016} and object detection~\cite{liuGenerativeModelingSmallData2019}.
Generative AI models such as Generative Adversarial Networks (GANs) and Fully Convolutional Neural Networks (FCNs) contain an `upscaling' mechanism to generate new data.
For example, generator modules in GANs contain specialized layers to upscale input feature maps, where the Transposed Convolution (TCONV) layer is the core of this `upscaling’ mechanism.
As the use of these types of models increases throughout different applications, there is an ever-growing need to make them more efficient on resource-constrained edge devices. 
However, this requires exploiting across-stack optimizations~\cite{gibson_dlas_2024} compared to executing them on devices with powerful CPUs/GPUs.
When compared to the traditional convolution operation, which has been extensively studied~\cite{mittalSurveyFPGAbasedAccelerators2020}, the complex computing properties of the TCONV operation, such as the overlapping sum problem~\cite{zhangDesignMethodologyEfficient2017a}, make it challenging to design accelerators that efficiently process TCONV, especially on edge devices with limited computational and memory capabilities.

Previous work has focused on improving upon different TCONV implementation methods on specialized accelerators.
For example, the Zero-Insertion~\cite{yuUniOPUFPGABasedUniform2020a} and Transforming Deconvolution to Convolution (TDC)~\cite{changEnergyEfficientFPGABasedDeconvolutional2020} methods have computational and transformation overheads, thus researchers have been focusing on the Input-Orientated-Mapping (IOM) method~\cite{yanGNAReconfigurableEfficient2018a} for TCONV.
IOM reduces the number of operations required to perform TCONV without requiring additional padding or transformations to inputs or weights.

\begin{figure}[!t]
    \centering
    \includegraphics[width=0.48\textwidth]{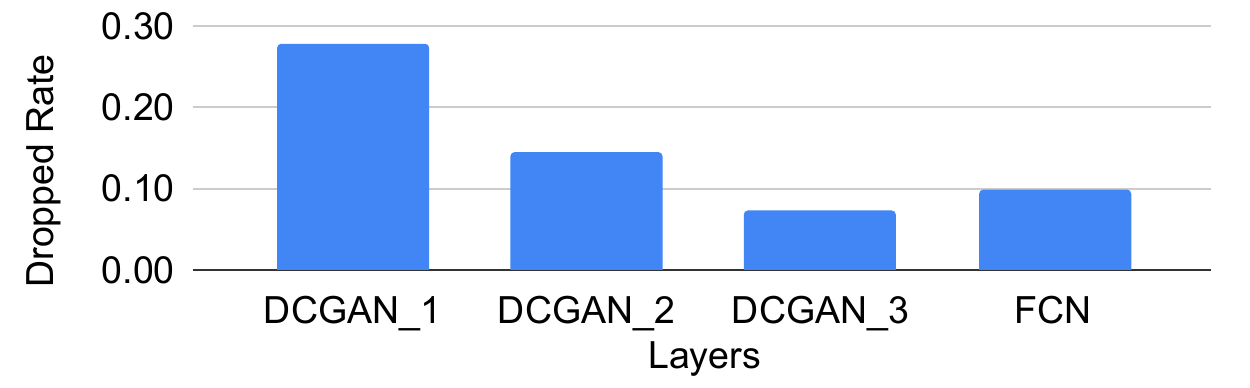}
    \caption[The \% of cropped outputs for various TCONV problems]{Percentage of cropped outputs for the various TCONV problems benchmarked in our evaluation.}\label{fig:drop_out}
\end{figure}

However, efficient execution of IOM on resource-constrained edge devices requires a hardware-software co-designed approach to ensure optimized tiling and offloading of the TCONV operation to the accelerator, while tackling three key interlinked problems effectively:
\begin{enumerate*}[label=(\roman*)]
\item storing intermediate/partial results;
\item processing overlapping sums;
\item handling cropped outputs.
\end{enumerate*}
First, partial results generated during TCONV should remain in on-chip memory to reduce the latency of sending data back to main memory.
This means that on a device with limited memory space, storage of results should be optimized so that it takes up minimal space.
Second, the overlapping sum problem occurs when multiple spatially separate dot product operations produce partial results of a single output value; these values must be coalesced into a single output value.
Since the spatial locality of the partial results corresponding to a single final output varies for each output and also between the TCONV problem dimensions, creating a specialized accelerator to handle the complex mapping of partial values to the output efficiently becomes challenging.
Finally, the standard IOM approach using Matrix Multiplication (MatMul) creates additional output data that needs to be cropped from the final results to maintain consistent dimensions across the model execution.
Therefore, the IOM approach not only leads to ineffectual computation, as the additional values are dropped later, but also to transformation overheads due to output cropping.
Figure~\ref{fig:drop_out} highlights the number of outputs dropped during the output cropping process across various TCONV-based GAN layers, which is proportional to the number of wasted operations.

To the best of our knowledge, existing accelerator solutions~\cite{maIntermediateCentricDataflowTransposed2023,zhangDesignMethodologyEfficient2017a,xuFCNEngineAcceleratingDeconvolutional2018,sestitoFPGADesignTransposed2023} for TCONV do not tackle these three problems efficiently on resource-constrained edge devices.
They especially neglect the issue of the ineffectual computations due to calculation of outputs that are cropped. 

To address this problem, in this paper we present \textbf{MM2IM}, a novel hardware-software co-designed accelerator for TCONV that merges \textbf{M}at\textbf{M}ul with col\textbf{2IM}~\cite{matlabdevsCol2imRearrangeMatrix}, a matrix transformation operation that rearranges data columns into blocks.
Our approach utilizes algorithmic optimizations along with specialized hardware modules to enable the IOM method on resource-constrained edge devices by efficiently handling the previous three key challenges, i.e., the overlapping sum problem, ineffectual computations due to cropped outputs, and tiling TCONV computations.
We develop our design using the SECDA-TFLite~\cite{harisSECDATFLiteToolkitEfficient2023} toolkit and evaluate its performance on an edge platform that includes an FPGA across various TCONV problems, including end-to-end execution of GAN models.
Furthermore, we compare the performance of MM2IM against similar TCONV accelerators for resource-constrained edge FPGAs, demonstrating superior throughput per DSP.
The contributions of this paper can be summarized as follows:

\begin{itemize}
    \item \textbf{MM2IM}: a new accelerator architecture that solves the three main TCONV problems and efficiently processes TCONV operations on resource-constrained edge devices using our custom IOM-based tiling strategy.

    \item \textbf{Two specialized hardware modules}: the TCONV mapping engine that generates compute and output maps on-the-fly for any TCONV problem configuration, and the processing module that efficiently computes and stores the required outputs while skipping ineffectual computations using the compute and output maps.
    
    \item \textbf{Integration and evaluation of MM2IM}: we integrate MM2IM within TFLite and run a range of experiments that compare MM2IM against our ARM Neon optimized CPU baseline and other FPGA-based resource-constrained accelerators for TCONV. 
    We obtain an average speedup of \SynthSpeedUp across \SynthCount TFLite TCONV problem configurations and, similarly, achieve up to \uptoModelSpeedUp speedup and \uptoModelEnergyreduction energy reduction across two GAN models while improving GOPs/DSP efficiency by at least \atleastGOPsDSPx over other accelerators.  
\end{itemize}

\section{Background}


\subsection{Transposed Convolution}\label{subsec:tconv}

Transposed Convolution (TCONV) is the key operation used within generative AI models to enable `upscaling' of input data. The TCONV parameters are defined as:
\begin{equation}\label{eq:tconv}
  out(O_h,O_w,O_c) = tconv(I_h,I_w,I_c,Ks,O_c,S)
\end{equation} 
where $I_h,I_w,I_c$ are the input height, width and channels, respectively; with kernel size $Ks$, output channels $O_c$ and stride $S$.
The output dimensions, height $O_h$ and width $O_w$, are defined as: $O_{h\\w}\mathord{=}S \times I_{h\\w}$.
When $Ks$ $>$ $S$, executing the direct TCONV operation requires the coalescing of partial outputs into the same final output due to striding; this coalescing is known as the overlapping sum problem~\cite{yanGNAReconfigurableEfficient2018a}.
Since the mapping of partial outputs to final outputs is problem dependent, the complexity of the output mapping increases.

Alternatively, there are three optimized methods for implementing TCONV: 
\begin{enumerate*}[label=(\roman*)]
  \item Zero-Insertion;
  \item Transforming Deconvolution to Convolution (TDC);
  \item Input-Oriented Mapping (IOM).
\end{enumerate*} 
Zero-Insertion resolves the overlapping sum problem by padding the input zeros, albeit with added compute, memory, and bandwidth overhead, approximately 75\%~\cite{xuFCNEngineAcceleratingDeconvolutional2018}. 
The TDC method transforms TCONV operations into Convolution operations by generating sub-filter kernels to avoid the overlapping problem, but this method requires additional hardware to process the sparse sub-filter efficiently~\cite{changEnergyEfficientFPGABasedDeconvolutional2020}.
For the IOM method, introduced within GNA~\cite{yanGNAReconfigurableEfficient2018a}, each activation is multiplied by the filters and then the overlapped partial results are summed to produce the final output.
The IOM method reduces the number of operations required to perform TCONV, as it does not require additional padding or transformation of inputs or weights.
The drawback of the IOM method is that it contains a significant amount (up to 28\% for DCGAN~\cite{radfordUnsupervisedRepresentationLearning2016}) of ineffectual computations due to cropped outputs.
Additionally the IOM method also requires an efficient hardware architecture to overcome the overlapping sums problem.


\subsection{Input-Oriented Mapping using MatMul and col2IM}

\begin{figure}[!t]
    \centering
    \includegraphics[width=0.48\textwidth]{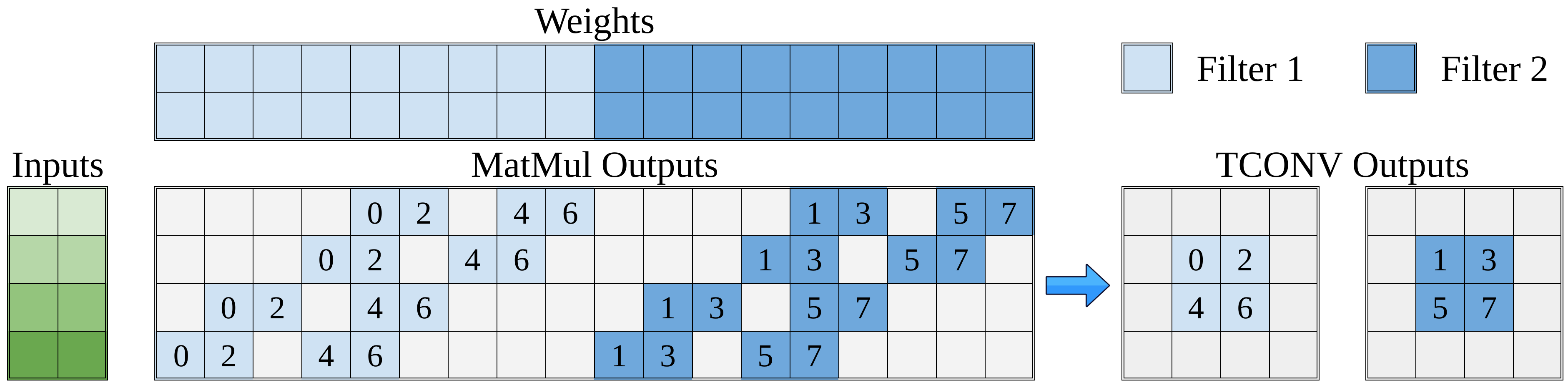}
    \caption{Example of TCONV with MatMul + col2IM.}
    \label{fig:tconv_example}
    \label{fig:data_layout}
\end{figure}

Here we discuss the IOM method in more detail. We express the IOM method in terms of the following operations: 
\begin{equation}\label{eq:iom}
  out(O_h,O_w,O_c) = col2im(mm(I,W_T),O_h,O_w,O_c)
\end{equation}
where $I(I_h,I_w,I_c)$ is the input data, $W(Ks,Ks,O_c,I_c)$ is the filter data, $mm$ is the matrix multiplication (MatMul) operation, and $col2im$~(column to image) is the operation used to convert the output of the MatMul to the final output.
Figure~\ref{fig:tconv_example} highlights the TCONV operation using the IOM method implemented for an example TCONV problem $tconv(2,2,2,3,2,1)$.
Translating the TCONV dimensions to MatMul dimensions, we define the dimensions as: $M = I_h * I_w$, $N = Ks^2 * O_c$, and the depth dimension $K = I_c$.
Hence, the partial output matrix can be represented by dimensions $M$ and $N$, i.e., the rows and columns of the MatMul operation; 
and the number of operations required by the IOM method is equivalent to $I_h * I_w * I_c^2 * Ks^2 * O_c$ or simply $M * N * K$.
Once the partial outputs are calculated, the final output is determined through the col2IM operation, which accumulates the partial outputs into the final TCONV outputs.

Note that the IOM method produces padded output feature maps, so the perimeter of the output feature maps are cropped, as shown by the gray squares in Figure~\ref{fig:tconv_example}.
Each partial output requires a dot product of the input row and filter column.
The partial output is then summed to produce the final output; hence, each gray square computed represents $K$ ineffectual computations.
These partial outputs must be stored within temporary output buffers until all the related partial outputs are calculated, since the standard IOM method with MM2IM does not consider output mapping during the matrix multiply operation.


\section{Efficient Transposed Convolution}


\subsection{Optimizing Input-Oriented Mapping}\label{subsec:iom}

To optimize TCONV using the IOM method we first elaborate on its main inefficiencies in detail and then discuss optimizations for better performance on resource-constrained edge devices.

\subsubsection{IOM Inefficient Computation}

The baseline IOM method has two inefficiencies: ineffectual computations and the storage of partial outputs.
The number of ineffectual computations, i.e., the dropped outputs $D_o$ per TCONV problem, can be statically determined using the col2IM algorithm~\cite{matlabdevsCol2imRearrangeMatrix}. 
Overall, for a given TCONV problem, the IOM efficiency can be determined by looking at the drop rate: $D_r = D_o / (M * N) $.
In the example in Figure~\ref{fig:tconv_example}, where $D_o=40$ and $M*N=72$, $D_r = 0.55$, hence $55\%$ of the computations is unnecessary.


\subsubsection{IOM Inefficient Storage}

In terms of storing partial outputs, we can calculate the wasted buffer space $W_s$ as the number of final outputs $F_{outs}$ minus the number of partial outputs $P_{outs}$, assuming that we do not skip ineffectual computations; 
where $F_{outs} = O_c * O_h * O_w$ and $P_{outs} = M * N$.
In an ideal scenario, we can completely skip storing partial outputs and simply accumulate them to the final output, thus improving buffer space efficiency by $P_{outs} / F_{outs}$.
In the case of the example in Figure~\ref{fig:tconv_example}, where $P_{outs} = 72$ and $F_{outs} = 32$, this would improve space efficiency by $2.25~\times$.
If we are also able to skip ineffectual computations, we improve the buffer space efficiency up to $9~\times$ for this example.

\subsubsection{Solving IOM Inefficiencies}

To solve the inefficiencies of IOM, we first define the \textit{output mapping} and the \textit{compute mapping}.
In Figure~\ref{fig:tconv_example}, each square in the \textit{MatMul Outputs} represents a partial TCONV output, and the values inside these squares represent the output index of the final TCONV outputs (shown on the right); this \textit{output mapping} is a function of $S$ and the input dimensions $(I_h, I_w)$.
For example, all the `0' index partial outputs are summed and stored in the `0' index of the final output feature maps.
Additionally, calculating the index map of the (light and dark) blue squares in the \emph{MatMul Outputs}, we derive the \textit{compute mapping} for the given TCONV problem, that is, the index map of partial outputs that are not dropped out via col2IM.

Therefore, our \textbf{first key insight} is that by using the \textit{output mapping} and the \textit{compute mapping}, we can solve the IOM inefficiencies and enable an efficient accelerator architecture that can:
\begin{enumerate*}[label=(\roman*)]
 \item Skip ineffectual computations of the dropped partial outputs (gray squares);
 \item Remove the need for storing partial sums in temporary memory and to be summed later;
 \item Map the outputs of the MatMul operation directly to the final output values.
\end{enumerate*}


\subsection{Acceleration Dataflow for Resource-Constrained Devices}

Data transfer between off-chip and on-chip memory can become a bottleneck, especially on resource-constrained edge devices.
Hence, we co-designed \textit{Tiled MM2IM}, a specialized tiling strategy for MM2IM that enables weight and output stationary dataflow minimizing data transfer redundancy, highlighted in Algorithm~\ref{alg:tiled_mm2im}.
Tiled MM2IM loads $filter\_step$ filters and produces the corresponding output channels within the outer loop.
The $filter\_step$ is determined by the number of processing modules within our MM2IM architecture (discussed in Section~\ref{sec:design}).
We dynamically load the input rows to calculate one output row per iteration within the inner loop; the $i\_end\_row$ array that holds the number of input rows required to compute the current output row is pre-calculated.

Therefore, our \textbf{second key insight} is that using the previous dataflow can increase/decrease hardware parallelism depending on the resource constraints by adjusting $filter\_step$.
Additionally, with \textit{Tiled MM2IM}, we preemptively calculate partial outputs for later output rows depending on the input rows being processed.

\begin{algorithm}[t]
\footnotesize
  \caption{Tiled MM2IM}\label{alg:tiled_mm2im}
  \DontPrintSemicolon
  \KwData{Initialize $filter\_step$, $i\_end\_row$}
  \ForEach{$c \gets 0$ \KwTo $O_c$ \textbf{by} $filter\_step$}{
    SendWeightFilters($c$, $filter\_step$)\;
    $starting \gets 0$\;
    \ForEach{$h \gets 0$ \KwTo $O_h$}{
      $rows\_to\_send \gets i\_end\_row[h] + 1 - starting$\;
      \If{$i\_end\_row[h] \neq starting - 1$}{
        SendInputRows($starting$, $rows\_to\_send$)\;
      }
      ComputeOutRow($h$, $c$, $filter\_step$)\;
      StoreOutRow($h$, $c$, $filter\_step$)\;
      $starting \gets i\_end\_row[h] + 1$\;
    }
  }
\end{algorithm}


\subsection{Performance Model}\label{subsec:performance_model}

We built an analytical model for our MM2IM architecture to estimate performance and guide further design choices.
Our performance model accounts for the problem size and the properties of our accelerator design to assess the overall performance.
Additionally, we combine accelerator analysis with data movement analysis to estimate the end-to-end performance for a given TCONV layer.
Next we provide an overview of our performance model.

First, we calculate problem-specific metrics such as the number of MAC (multiply and accumulate) operations and the number of cropped MatMul outputs for the given TCONV problem.
Then, we calculate the accelerator processing time, finding the processing time for each Processing Module (PM) and its components, the Compute Unit (CU) and the Accumulation Unit (AU), which are all discussed in detail in Section~\ref{subsec:pm}: 
\begin{equation}
    T_{PM} = T_{CU\_compute} + T_{CU\_load} + T_{CU\_store} + T_{AU}
\end{equation}
Then, we calculate the data transfer time required between main memory and the accelerator:
%
\begin{equation}
    T_{Data} = (W_{size} + I_{size} + O_{size} + OMap_{size}) * BW
\end{equation} 
Finally, we calculate the end-to-end latency for the TCONV operation, combining the two factors: $T_{total} \mathord{=} T_{PM} + T_{Data}$.

Therefore, our \textbf{third key insight}, through performance modeling, is that up to $35\%$ of the end-to-end latency ($T_{total}$) for any given TCONV problem was due to transferring output mapping data between the main memory and the accelerator.
Hence, we developed the MM2IM mapper, a hardware module that completely removes the need for output mapping data transfers, discussed in Section~\ref{subsec:mm2im_mapper}.

\section{MM2IM Accelerator Architecture}\label{sec:design}

\begin{figure}[!t]
  \centering
  \includegraphics[width=0.44\textwidth]{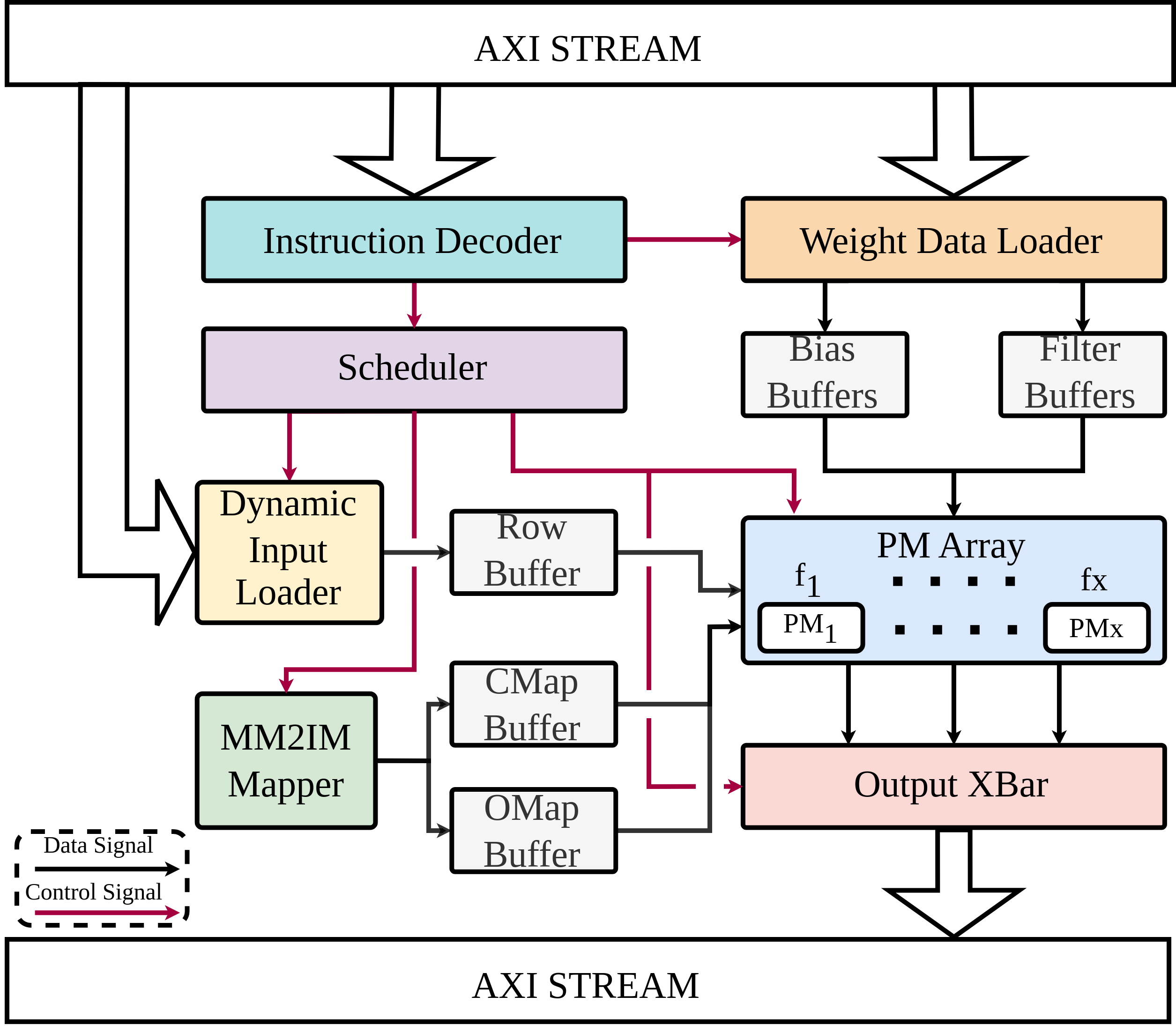}
  \caption{MM2IM Accelerator Architecture. The accelerator is connected to main memory via AXI-Stream buses, which are used to receive instructions and send/receive data.}
  \label{fig:acc_fig}
\end{figure}

Figure~\ref{fig:acc_fig} overviews MM2IM, our proposed stream-based and scalable accelerator architecture, which utilizes simple instructions to configure, load data and execute TCONV operations.
These instructions enable MM2IM to dynamically tile TCONV operations using Algorithm~\ref{alg:tiled_mm2im}.
MM2IM exploits two dimensions of parallelism at the \emph{Processing Module} (PM) level, splitting $O_c$ (output channels) computations across the $X$ number of PMs (used for $filter\_steps$ in Algorithm~\ref{alg:tiled_mm2im}) and unrolling $I_c$ (input channels) within the compute units with an unrolling factor of \textit{UF}.
Note that for our instantiation, we have set $X$$=$$8$ and \textit{UF}$=$$16$; these parameters could be scaled to meet performance demands and resource constraints.
The following sections discuss the key components of MM2IM.

\subsection{Instruction Decoder}\label{subsec:id}

The \emph{instruction decoder} allows for the reconfiguration of the accelerator to enable the execution of various TCONV layer configurations.
It decodes instructions and sends control signals to the \emph{\emph{Scheduler}} and the \emph{weight data loader}.
Table~\ref{tab:opcodes} shows the micro-ISA opcode set for the accelerator and a brief description of each instruction.
These instructions are generated and sent to the accelerator by the host-side driver code during the execution of a given TCONV layer.
Note that some opcodes, for example `0x01', are immediately followed by operand data which the accelerator expects once the instruction is decoded; this enables dynamic reconfiguration of the accelerator or loading new sets of data into the accelerator buffers.

\begin{table} [t]
  \centering
  \caption{Micro-ISA Opcode Set.}
  \label{tab:opcodes}
  \begin{tabular}{|c|c|}
  \hline
  \textbf{Opcode} & \textbf{Description}   \\ \hline
  0x01            & Configure TCONV (sets configuration registers)        \\ \hline
  0x02            & Loads Bias and Filter (activates Weight Data Loader)            \\ \hline
  0x04            & Load Input (activates Dynamic Input Loader)             \\ \hline
  0x08            & Schedule TCONV (activates Scheduler)         \\ \hline
  0x10            & Store Output (activates Output Crossbar)           \\ \hline
  \end{tabular}
\end{table}

\subsection{Scheduler}\label{subsec:scheduler}

The \emph{Scheduler} is the main control unit within the accelerator.
Once activated, it orchestrates the execution of an entire TCONV layer.
First it activates the \emph{MM2IM Mapper} alongside the \emph{Dynamic Input Loader}, and then it activates the array of \emph{Processing Modules} (PMs) to execute the operations required by the TCONV layer.
Additionally, the \emph{Scheduler} continuously monitors the \emph{Instruction Decoder} for new instructions to either load the next row of input data to the \emph{Row Buffer} or to send back the output data to main memory by activating the \emph{Output Crossbar}.
The \emph{Scheduler} has fine-grained control over the PMs, allowing them to be turned on or off as needed.


\subsection{Data I/O}\label{subsec:data_loader}

There are two data loaders, the \emph{Weight Data Loader} and the \emph{Dynamic Input Loader}.
The \emph{Weight Data Loader} loads batches of filter and bias data from main memory (via AXI-Stream) to their respective buffers.
Once a batch is loaded into the buffers, the \emph{Scheduler} allocates the filter and the corresponding bias data across the PMs.
The \emph{Dynamic Input Loader} loads new rows of inputs data dynamically to store within the \emph{Row Buffer}, which at the request of the \emph{Scheduler} broadcasts the new row of input data to all the PMs via dedicated FIFOs.

The \emph{Output Crossbar} is an interface module which combines the output streams from each of the PMs, and at the request of the \emph{Scheduler} it sends the output data back to main memory.
Note that the \emph{Scheduler} sends a store request when the `Store Output' instruction is received by the accelerator.





\subsection{Processing Module Array}\label{subsec:pm}

\begin{figure}[!t]
  \centering
  \includegraphics[width=0.45\textwidth]{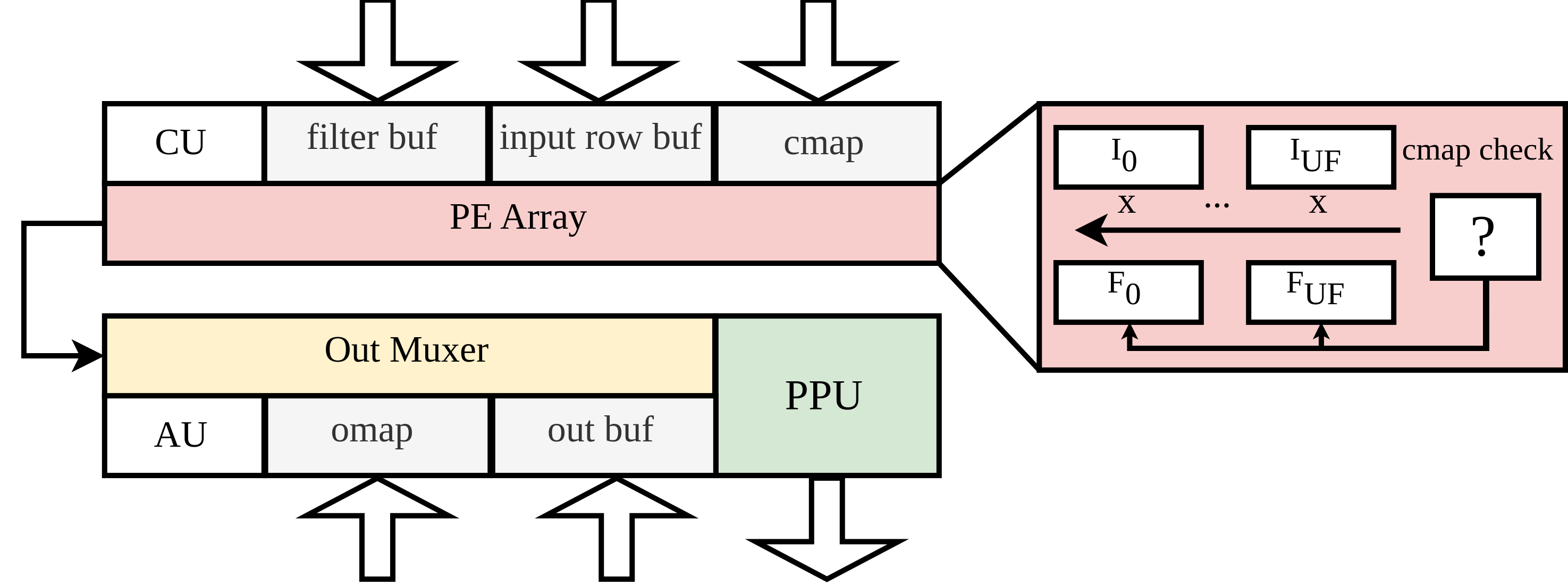}
  \caption{Processing Module architecture with a detailed view of the PE array. The wide white arrows represent data movement from and to the rest of the accelerator.}
  \label{fig:pm_fig}
\end{figure}

The \emph{Processing Module} (PM) array is the accelerator's computational core. 
It consists of $X$ PMs that can be individually configured and utilized by the \emph{Scheduler} to ensure efficient processing of TCONV.
Each PM contains an accumulation and a compute unit connected via a FIFO stream.
Figure~\ref{fig:pm_fig} provides a detailed view of the PM architecture with a fine-grained view of the PE array;
note that the wide white arrows represent data movement between the rest of the accelerator and the \emph{Processing Module}.

For each TCONV layer, $X$ filters are partitioned along the PMs.
Once all the PMs load their respective filters, rows of input data are streamed to all the PMs.
Additionally, the PMs receive the compute map (\textit{cmap}) and output map (\textit{omap}) from the \textit{MM2IM Mapper} (described in Section~\ref{subsec:iom}).

\textbf{Compute Unit:}
Due to the complex computing nature of TCONV, the Compute Units (CUs) contain additional logic to ensure that ineffectual dot product computations are skipped.
This additional logic, the `cmap check' within the \emph{PE Array}, takes the \textit{cmap}, the input row, and the filter data, and computes the dot product of the \textit{selected} input row and filter column.

The partial results are then streamed into the accumulation unit for further processing.
Additionally, CUs are scalable and the unrolling factor (UF) defines the number of MACs within the PE array per CU.
The UF is used to tile the $I_c$ dimension of the given TCONV layer.
Hence, to execute dot-product from $I_c$, the PE array will take $I_c/UF$ number of cycles.
Increasing \textit{UF} will directly increase the number of MACs per cycle per CU while increasing the hardware resources required.

\textbf{Accumulation Unit:}
The partial sums calculated by the CUs are stored within the \textit{output buffers} in the correct output indices; the \textit{Out Muxer} ensures this by using the \textit{omap}.
Subsequent partial sums for the same output accumulate with existing results, avoiding the need for extra buffer space. 
Once the output is fully calculated for an entire output row, the post-processing unit (\textit{PPU}) processes the row.
The \textit{PPU} is a specialized processing engine used to perform the post-processing steps required by a given DNN model and then send the output data to the \emph{Output Crossbar}.


\subsection{MM2IM Mapper}\label{subsec:mm2im_mapper}

The \textit{MM2IM Mapper} is a key component of our accelerator architecture, as shown in Algorithm~\ref{alg:mm2im_mapper}.
It generates the \textit{cmap} and \textit{omap} corresponding to the row of partial results (i.e., the output row of MatMul) and streams them to the PMs. 
Note that it takes the current $row_{id}$ and the number of rows as parameters to ensure that cmap/omap are generated only for the required rows.
This allows the MM2IM Mapper to support partitioning of the data in a tiled manner, since the $row_{id}$ can be initialized to the starting row of the output tile instead of the starting row of the output matrix.
Overall, the MM2IM Mapper generates the compute and output mappings only once per row, and each map is broadcast to all PMs, thus saving hardware resources and additional computational overhead.
Note that the MM2IM Mapper is configured dynamically through the `0x01' opcode, and can support any shape of TCONV layers.

\begin{algorithm}[!t]
\footnotesize
  \caption{MM2IM Mapper}\label{alg:mm2im_mapper}
      \DontPrintSemicolon
          $row_{id} \gets${load($row_{id}$)}, $row_{width} \gets${load($row_{width}$)}\;
          \ForEach{$r$ in $MM_{rows}$}{
              $h_{pad} = -padding_{top} + (S * (row_{id} ~\%~ row_{width}))$\;
              $w_{pad} = -padding_{left} + (S * (row_{id} \div row_{width}))$\;
              $im_{dex}= h_{pad}  * O_{w} + w_{pad}$\;
              $col = 0$\;
              \ForEach{$ih$ in $Ks$}{
                  \ForEach{$iw$ in $Ks$}{
                      \uIf{($ih + h_{pad} >= 0 ~\&\&~ ih + h_{pad} < O_h$ \\ $~\&\&~ iw + w_{pad} >= 0 ~\&\&~
                    iw + w_{pad} < O_w$)} {
                          $PMs\_cmap.broadcast\_write(col)$\;
                          $PMs\_omap.broadcast\_write(im_{dex})$\;
                      }
                      $col++$, $im_{dex}++$\;
                  }
                  $im_{dex} += O_{w} - Ks$
              }
              $row_{id}++$
          }
\end{algorithm}


\subsection{Computational Flow}\label{subsec:com_flow}

\begin{figure*}[!t]
  \centering
  \includegraphics[width=0.97\textwidth]{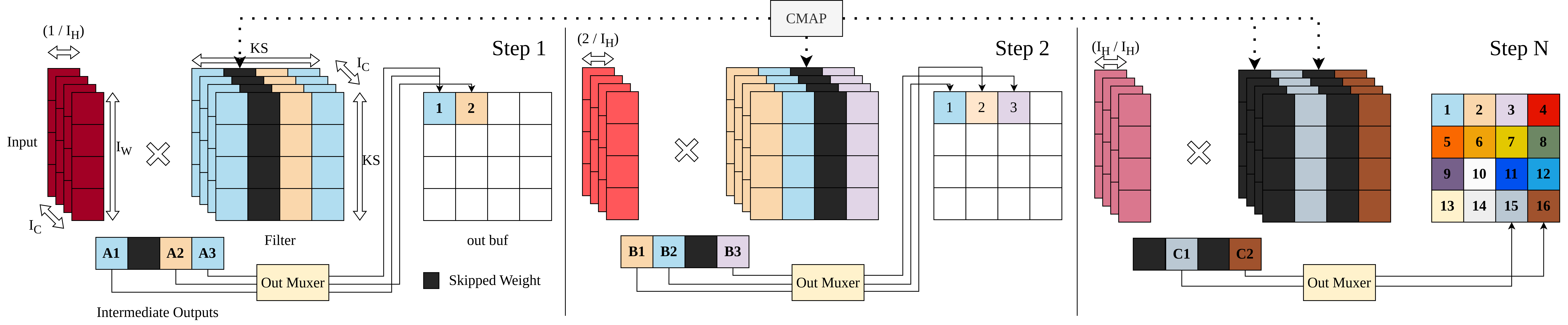}
  \caption{Computational flow within each PM in terms of MM2IM computations.}
  \label{fig:com_flow}
\end{figure*}

Finally, we discuss in more detail the flow of computation within each PM and how each PM is able to calculate an output channel across multiple steps.

Figure~\ref{fig:com_flow} provides an example of an input feature map being processed through a single filter to produce a complete output channel over $N$ steps ($N = I_H$); this happens within each PM. 
The inputs on the left of each step represent the input row (presented as a column in Figure~\ref{fig:com_flow}), which will be processed by all PMs during that step.
Each input row contains $I_w * I_c$ data elements; this data is fetched from the `Row Buffer' after each step, as seen in Figure~\ref{fig:acc_fig}.
Before starting the computations, each PM is preloaded with a single filter (shown in the middle); the filter contains $Ks *Ks*I_c$ data elements, which are stored in the PM's local buffer.
Note that the color of filter columns represents the intermediate outputs they produce within that step, and the weight data remains the same as the filter does not change between steps.

During each step, the PM array performs a dot product between the input row and all the non-skipped weight columns.
Some weight columns are skipped to ensure no ineffectual computation takes place; the indexes for the skipped weight columns are stored with the `CMAP'.
The intermediate outputs of dot product operations for each step are shown in Figure~\ref{fig:com_flow}. 
For `Step 1', intermediate outputs \emph{A1}, \emph{A2} and \emph{A3} are computed.
Once computed, the intermediate outputs are sent to the `Out Muxer' to map and accumulate within the correct output index of the local `out\_buf.' 
We see that, \emph{A1} and \emph{A3} are mapped to output \emph{1} whereas \emph{A2} is mapped to output \emph{2}.
In `Step 2', \emph{B2} is accumulated with the previous results of output \emph{1}, \emph{B1} is accumulated within output \emph{2}, and \emph{B3} is stored in output \emph{3}.
After $N$ steps, the entire output feature map is calculated.
One key feature of the MM2IM architecture is that the output data stored in `out\_buf' is sent back to main memory as soon as a whole row of outputs is completely accumulated, which enables us to reduce the size of `out\_buf.'

\section{Evaluation}


\subsection{Experimental setup}

To design, validate and evaluate our MM2IM accelerator, we utilized the SECDA methodology~\cite{harisSECDAEfficientHardware2021} for quick design and integration of its architecture.
Since our focus is on edge-based inference, we used the PYNQ-Z1 FPGA board, which includes a dual-core ARM Cortex-A9 CPU and an edge FPGA.
Additionally, we used TensorFlow Lite (TFLite) along with the SECDA-TFLite toolkit~\cite{harisSECDATFLiteToolkitEfficient2023} for the accelerator integration.

More specifically, we used SECDA-TFLite to develop a custom MM2IM delegate\footnote{A TFLite delegate is a hardware-software backend for DNN operations.} for TFLite; this custom delegate first selects all TCONV layers within the target TFLite model for offloading to our accelerator.
During inference, the selected layers will be processed by our MM2IM delegate, which offloads all TCONV-related metadata and pointers to the MM2IM driver code.
Then, the host driver orchestrates the tiling strategy for accelerating TCONV as described in Algorithm~\ref{alg:tiled_mm2im}, offloading the relevant input and weight data as required by the accelerator to calculate the output feature maps for the layer.
As the accelerator finishes calculating each output feature map, the accelerator sends back the output data to be stored in the TFLite allocated tensor in main memory.


\subsection{Synthetic benchmarks}

First, we evaluate the performance of our MM2IM accelerator across varying sets of TCONV problems.
Using the benchmarking suite available within SECDA-TFLite~\cite{harisSECDATFLiteToolkitEfficient2023}, we generate single-layer TCONV models and benchmark the performance of MM2IM across \SynthCount TCONV permutations.
We permuted the TCONV parameters with the following values that occur commonly in TCONV models:
\begin{enumerate*}[label=\roman*)]
 \item $O_c \mathord{=} [16,32,64]$;
 \item $Ks \mathord{=} [3, 5, 7]$;
 \item $I_h \mathord{=} [7, 9, 11]$;
 \item $I_c \mathord{=} [32, 64, 128, 256]$;
 \item $S \mathord{=} [1, 2]$.
\end{enumerate*}
We discussed these parameters in Section~\ref{subsec:tconv}.

\begin{figure}[!t]
    \centering
    \includegraphics[width=0.48\textwidth]{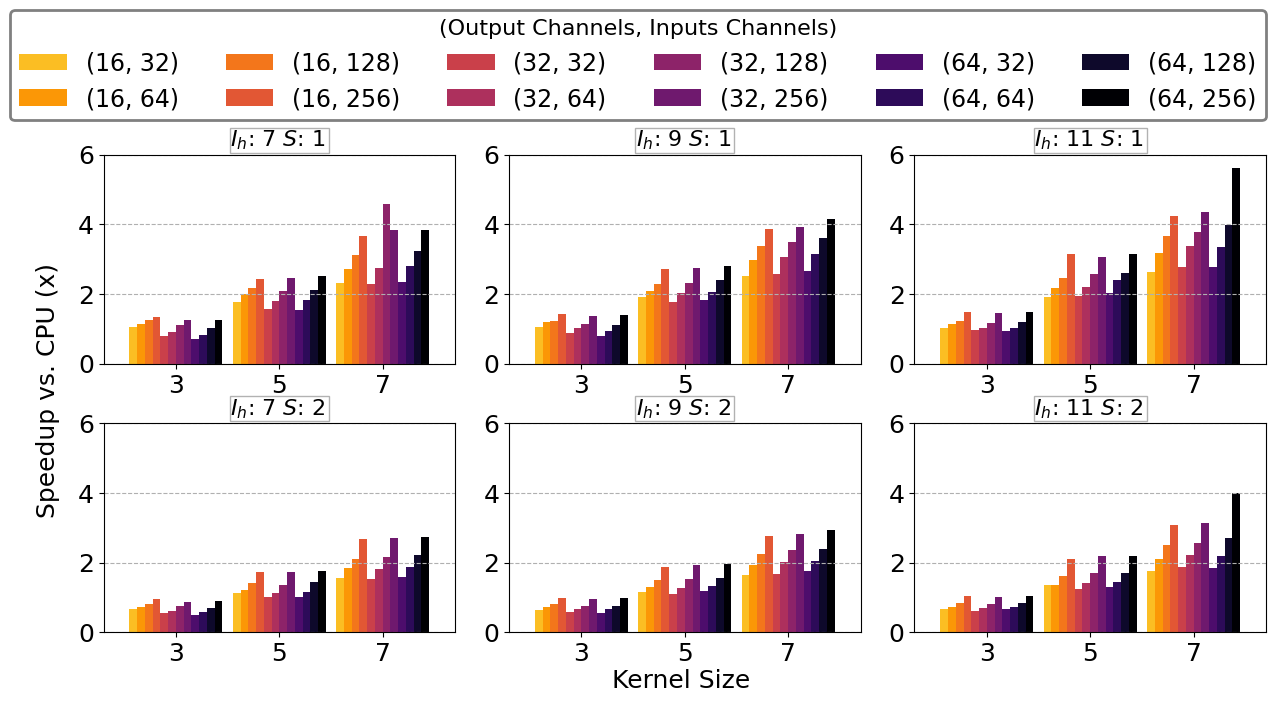}
    \caption{MM2IM speedup normalised to CPU execution time across various TCONV problems.}
    \label{fig:synth_fig}
\end{figure}

Figure~\ref{fig:synth_fig} shows the normalized speedup against dual-thread CPU 8-bit baseline (with NEON-vector instructions enabled) of the same problems on the PYNQ-Z1 board.
We group similar problems for ease of visualization of the results.
This set of experiments was designed to help us understand our accelerator's performance on a wide variety of TCONV problems, more specifically, to understand the dynamics between TCONV dimensions and accelerator performance.
The key takeaways from these experiments are the following:
\begin{enumerate*}[label=\roman*)]
    \item On average, MM2IM achieves a \SynthSpeedUp speedup against the dual-thread CPU;
    
    \item The larger the $I_c$ dimension, the greater the speedup - detailed profiling revealed that since the $I_c$ is not tiled and is processed together without off-chip memory access, the PE array utilization can remain higher;
    
    \item Similarly, higher $I_h$, $I_w$, $Ks$ dimensions also achieve greater speedup - this is due to higher data-reuse of the filter ($Ks^2$) data within each PM when the $I_h$ and $I_w$ dimension is larger;
    
    \item As expected, increasing the $O_c$ dimension yields a relatively smaller role in performance uplift; this is because MM2IM is tiled in the $O_c$ dimension, and hence the speedup due to higher $O_c$ is capped by the number of PMs. 
    
    \item Higher stride values result in lower speedup (on average 54\%), as expected, due to less cropped outputs.
\end{enumerate*}

\begin{figure}[!t]
    \centering
    \includegraphics[width=0.48\textwidth]{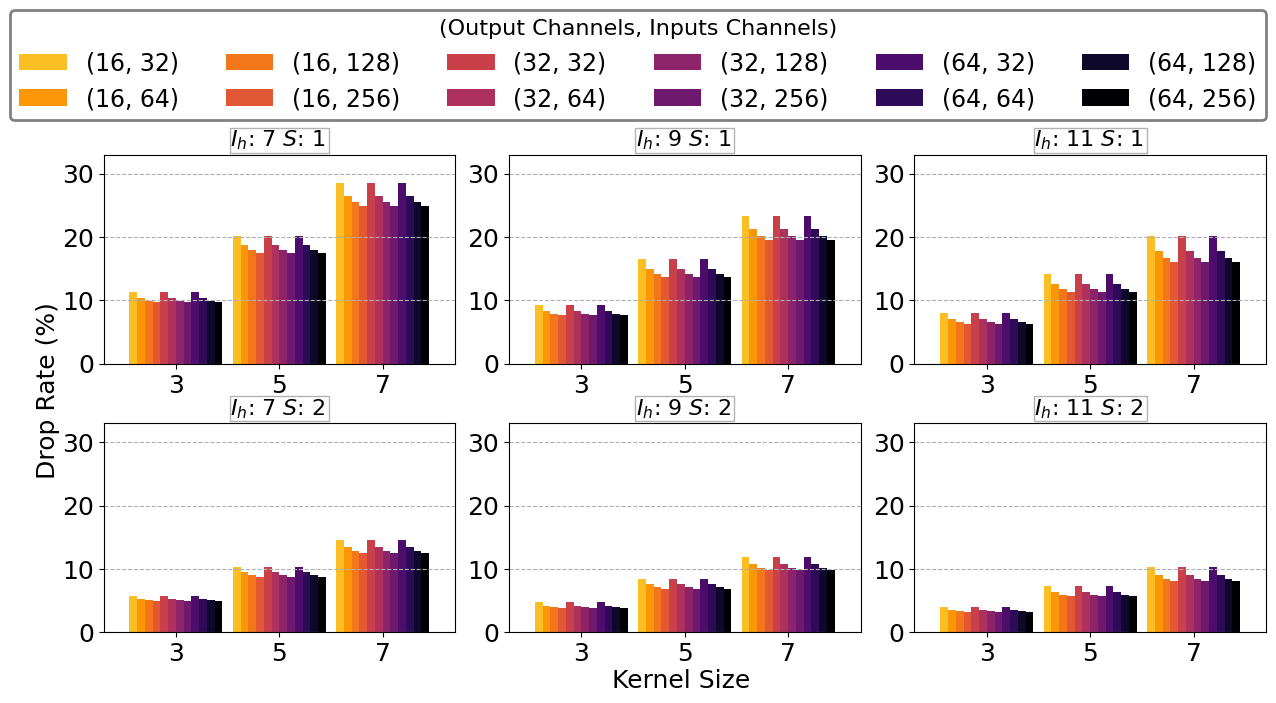}
    \caption[The \% of cropped outputs for various TCONV problems]{Percentage of cropped outputs for the various TCONV problems benchmarked in Figure~\ref{fig:synth_fig}.}\label{fig:droprate_fig}
\end{figure}

To highlight the impact of cropped outputs on the speedup, we generated Figure~\ref{fig:droprate_fig}, which highlights the \% of cropped outputs (`drop rate') for the various TCONV problems benchmarked within Figure~\ref{fig:synth_fig}.
The drop rate is calculated as the ratio of cropped outputs to the total number of outputs.
Looking at Figure~\ref{fig:droprate_fig}, we can see that increasing $Ks$ results in higher drop rates, while higher $I_h$ and $S$ result in lower drop rates.
Comparing the drop rate to the speedup, we can conclude:
\begin{enumerate*}[label=\roman*)]
    \item Increased kernel size results in higher drop rates and greater speedup;
    \item Increased stride results in lower drop rates and speedup, as expected;
    \item Decreased drop rate with increased $I_h$ does not hamper speedup.
\end{enumerate*}
We theorize that this is due to the increased utilization of the processing modules, as more computation is required for the larger input height and width.

\subsection{TCONV Model Layer Evaluation}

We performed an extensive evaluation across specific TCONV layers commonly found in popular generative models~\cite{radfordUnsupervisedRepresentationLearning2016,longFullyConvolutionalNetworks2015,johnsonPerceptualLossesRealTime2016,dongAcceleratingSuperResolutionConvolutional2016}.
Table~\ref{tab:model_layer_eval} contains the specific layer details, the performance of our accelerator compared to the PYNQ CPU's single-threaded execution, overall throughput, and energy efficiency.
On average, we achieve a $2.8\times$ speedup compared to the CPU implementation while achieving an average throughput of $5.5$ GOPs, with an average power to performance ratio of $14.9$ GOPs/W.
As realized in the synthetic benchmark, the MM2IM accelerator takes advantage of the larger $I_c$ dimension seen within the DCGAN~\cite{radfordUnsupervisedRepresentationLearning2016} layers, achieving up to a $4.2\times$ speedup in some layers.


\begin{table*}[t]
\centering
\caption{\label{tab:model_layer_eval} Performance Evaluation on Generative Model Layers.}
\begin{tabular}{|c|c|c|c|c|c|c|c|c|c|c|}
        \hline
        \textbf{Model} & \textbf{OC} & \textbf{KS} & \textbf{IH/IW} & \textbf{IC} & \textbf{OPs} & \textbf{Latency (ms)} & \textbf{CPU (ms)} & \textbf{Speedup (vs CPU)} & \textbf{GOPs} & \textbf{GOPs/W} \\
        \hline
        DCGAN\_1 & 512 & 5 & 4 & 1024 & 420M & 46.26 & 166.56 & 3.60 & 9.07 & 15.64 \\
        \hline
        DCGAN\_2 & 256 & 5 & 8 & 512 & 420M & 33.97 & 141.05 & 4.15 & 12.35 & 15.03 \\
        \hline
        DCGAN\_3 & 128 & 5 & 16 & 256 & 420M & 35.86 & 149.70 & 4.17 & 11.70 & 14.92 \\
        \hline
        DCGAN\_4 & 3 & 5 & 32 & 128 & 20M & 4.67 & 10.71 & 2.29 & 4.21 & 0.87 \\
        \hline
        FCN & 21 & 4 & 1 & 21 & 14K & 0.22 & 0.22 & 1.00 & 0.06 & 0.01 \\
        \hline
        StyleTransfer\_1 & 64 & 3 & 64 & 128 & 604M & 164.62 & 304.48 & 1.85 & 3.67 & 23.22 \\
        \hline
        StyleTransfer\_2 & 32 & 3 & 128 & 64 & 604M & 282.83 & 460.23 & 1.63 & 2.14 & 23.65 \\
        \hline
        StyleTransfer\_3 & 3 & 9 & 256 & 32 & 1020M & 264.27 & 1045.36 & 3.96 & 3.86 & 40.49 \\
        \hline
        FSRCNN & 2 & 9 & 32 & 32 & 11M & 5.21 & 12.47 & 2.39 & 2.04 & 0.51 \\
        \hline
    \end{tabular}
\end{table*}


\subsection{TCONV Accelerators Comparison}


\begin{table*}[h]
    \centering
    \renewcommand{\arraystretch}{1.2}
    \setlength{\tabcolsep}{5pt}
    \caption{\label{tab:comp}Comparison with state-of-the-art TCONV accelerators.}   
    \begin{tabular}{|c|c|c|c|c|c|}
        \hline
        \textbf{Source} & \textbf{\cite{zhangDesignMethodologyEfficient2017a}} & \textbf{\cite{liuOptimizingCNNbasedSegmentation2018a}} & \textbf{\cite{diExploringEfficientAcceleration2020}} & \textbf{\cite{changEnergyEfficientFPGABasedDeconvolutional2020}} & \textbf{Ours} \\
        \hline
        FPGA & ZYNQ 7Z020 & ZC706 XC7Z045 & ZC706 XC7Z045 & Kintex-7 XC7K410T & PYNQ Z1 \\
        \hline
        Frequency (MHz) & 100 & 200 & 167 & 130 & 200 \\
        \hline
        Precision & 12-bit & 16-bit & 16-bit & 13-bit & 8-bit \\
        \hline
        DSP Usage & 209 & 640 & 603 & 1512 & 49 (22\%) \\
        \hline
        LUTs Usage & 25K & 85K & 196K & 167K & 42K (79\%) \\
        \hline
        FFs Usage & 30K & 110K & -- & 158K & 49K (46\%) \\
        \hline
        BRAM Usage & 48\% & 67\% & 57\% & 24\% & 99\% \\
        \hline
        BRAM Space & 2.4MB & 12.8MB & 10.9MB & 6.7MB & 4.9MB \\
        \hline
        Perf. (GOPS) & 2.6 & 29 & 236.9 & 2691 & 23.0 \\
        \hline
        Perf. (GOPS/DSP) & 0.01 & 0.05 & 0.39 & 1.78 & 3.51 \\
        \hline
    \end{tabular}
\end{table*}

We now compare the performance improvement of MM2IM against other TCONV accelerators for resource-constrained edge FPGAs; Table~\ref{tab:comp} gives key details about the related works and summaries their performance.
Note that we do not compare to ASIC designs, as our work focuses on FPGA-based accelerator architecture for TCONV and direct comparisons to ASIC-based designs would not be fair, as translating area to FPGA resources is non-trivial.
In Table~\ref{tab:comp} we also present the best reported performance within each work.
While MM2IM does not achieve the highest throughput in terms of GOPs, we are able to outperform all the related works in-terms of GOPs/DSP performance, achieving a $2\times$ increase compared to the next best work~\cite{changEnergyEfficientFPGABasedDeconvolutional2020}, which utilizes the TDC method, while reducing LUT usage by $4\times$. 
The GOPs/DSP metric is more relevant than GOPs, as it takes into account the scale of the FPGA to ensure a fair comparison between accelerators.

The accelerators proposed by Liu et al.~\cite{liuOptimizingCNNbasedSegmentation2018a} and Zhang et al.~\cite{zhangDesignMethodologyEfficient2017a} serve as good points of comparison, as they target similar resource-constrained devices compared to our PYNQ-Z1.
Comparing these works, we outperform Zhang et al. by $8.8\times$ in terms of GOPs, while improving in DSP efficiency by $77\times$ compared to Liu et al.~\cite{liuOptimizingCNNbasedSegmentation2018a}.



\subsection{End-to-end GAN model evaluation}

\begin{table}[t]
\caption{\label{tab:e2e_res} End-to-end model inference performance of MM2IM against CPU-only inference dual-threaded execution.}
\centering
\fontsize{6.3}{9}\selectfont
\begin{tabular}{|c|c|cc|cc|cc|}
\hline
\textbf{Model} & \textbf{Configuration} & \multicolumn{2}{c|}{\textbf{TCONV (ms)}} & \multicolumn{2}{c|}{\textbf{Overall (ms)}} & \multicolumn{2}{c|}{\textbf{Energy (J/pic)}} \\ 
\hline
\multirow{4}{*}{DCGAN\footnotemark}
            & CPU 1T            & 38   & 1.0x  & 49   & 1.0x  & 7.9  & 1.0x  \\ \cline{2-8} 
            & ACC + CPU 1T      & 15   & 2.4x  & 21   & 2.3x  & 4.3  & 1.8x  \\ \cline{2-8} 
            & CPU 2T            & 24   & 1.6x  & 28   & 1.7x  & 6.5  & 1.2x  \\ \cline{2-8} 
            & ACC + CPU 2T      & 16   & 2.4x  & 20   & 2.4x  & 4.3  & 1.8x  \\ \hline  
\multirow{4}{*}{pix2pix} 
            & CPU 1T            & 2737  & 1.0x  & 5238   & 1.0x  & 9.8  & 1.0x  \\ \cline{2-8} 
            & ACC + CPU 1T      & 922   & 3.0x  & 3360   & 1.6x  & 7.9  & 1.2x  \\ \cline{2-8} 
            & CPU 2T            & 1532  & 1.8x  & 2886   & 1.8x  & 5.9  & 1.7x  \\ \cline{2-8} 
            & ACC + CPU 2T      & 926   & 3.0x  & 2266   & 2.3x  & 6.2  & 1.6x  \\ \hline  
\end{tabular}
\end{table}

To demonstrate the capability of our accelerator to accelerate end-to-end DNN execution, we evaluated MM2IM on two popular GAN models, DCGAN~\cite{radfordUnsupervisedRepresentationLearning2016} and pix2pix~\cite{isolaImagetoImageTranslationConditional2018} with end-to-end TFLite inference.
As we use unmodified TFLite models, we omit accuracy as it is unchanged with our accelerator; additionally, to validate correctness, we ensured that the accelerator output matches the CPU baseline output.

We accelerate the TCONV layers and the post-layer quantization using our MM2IM design.
The rest of the layers are executed on the board's CPU.
Table~\ref{tab:e2e_res} highlights the performance in terms of latency and power of MM2IM and CPU-only inference across the GAN models.

We achieve a latency improvement of \dualModelSpeedUp and an energy reduction of \dualModelEnergyreduction.
Note that since these GAN  models contain different types of layers, the potential end-to-end performance improvement with our accelerator is limited to the TCONV layers within the model.
For TCONV-only layers within the models, we achieve an average latency speedup of \dualModelTCONVSpeedUp compared with the dual-threaded MM2IM execution.


\subsection{Performance Model Validation}

As mentioned in Section~\ref{subsec:performance_model}, we used our performance model to estimate and guide the design choices of our MM2IM accelerator.
To validate our performance model, we compare its expected performance to our accelerator's actual performance.
On average, the model estimates the actual performance within $10\%$ of our MM2IM accelerator. 
Applying the TCONV decoder optimization to our performance model predicts the expected performance improvement within $1\%$ deviation of the actual performance improvement that the optimization provides.
This demonstrates the utility of our performance model in guiding design choices through estimated performance improvements per proposed optimizations.
Using our performance model helped us identify bottlenecks and solve the output mapping problems more efficiently using the MM2IM Mapper module.

\section{Related Work}


Various works with different accelerator architectures and optimizations have been proposed to accelerate TCONV operations, employing methods such as TDC~\cite{changEnergyEfficientFPGABasedDeconvolutional2020}, Winograd-Transformed Transposed Convolution~\cite{diExploringEfficientAcceleration2020,changDesignMethodologyEfficient2020}, and the Zero-Insert TCONV method~\cite{yuUniOPUFPGABasedUniform2020a}.
However, these approaches face computational overheads due to algorithmic limitations inherent to their respective methods.
~\footnotetext{We use the Tensorflow-defined version of DCGAN: \url{https://www.tensorflow.org/tutorials/generative/dcgan}}

Some works have proposed implementing the TCONV operation using the IOM method on FPGAs.
For example, Ma et al.~\cite{maIntermediateCentricDataflowTransposed2023} exploit the intermediate-centric dataflow, a variation on the IOM method, but their accelerator only supports fixed dimensions for given problems.
Similarly, the initial work by Sestito et al.~\cite{sestitoHighLevelSynthesisHardware2022} exploits High-Level Synthesis (HLS) to create `Deconvolution Engines' that can solve fixed dimension TCONV problems efficiently.
These engines require additional buffers to store the intermediate results until the entire filter is processed, after which the intermediate results are summed together to produce the final output feature map.
Additionally, this work does not handle the ineffectual computation preemptively, hence performing more MACs than required and also having to post-process the result to achieve the cropped results.
The later HLS template-based approach proposed by Sestito et al.~\cite{sestitoFPGADesignTransposed2023} suffers from the same constraints, and although they can adjust their accelerator to different problem sizes, this requires re-synthesis and re-mapping of the accelerator.
In both works, the computation engine processes a tile of $I_c$ dimension at a time, unlike our design, which processes the entire $I_c$ dimension, keeping the dataflow output stationary, allowing larger TCONV problems to be processed on smaller FPGAs without requiring slow off-chip memory access.
Zhang et al.~\cite{zhangDesignMethodologyEfficient2017a} proposed an output-oriented design that solves the overlapping sum problem for edge devices but introduces hardware complexity, degrading the accelerator's performance.

Other works such as GNA~\cite{yanGNAReconfigurableEfficient2018a} and FCN-Engine~\cite{xuFCNEngineAcceleratingDeconvolutional2018} exploit the IOM method with ASIC designs, but similar to all previously mentioned IOM-based approaches they do not consider the cropped outputs. 
Finally, there are several works~\cite{wuEfficientFPGABasedDilated2024,maIntermediateCentricDataflowTransposed2023} which take advantage of large-scale FPGAs to efficiently map TCONV problems fully within the FPGA fabric without requiring slow off-chip memory access.
While effective at tackling TCONV on larger FPGAs, they do not consider the limitations of TCONV within resources-constrained FPGAs.


\section{Conclusion}

We proposed MM2IM, a novel hardware architecture to efficiently accelerate TCONV operations on resource-constrained edge devices with FPGAs.
Our efficient hardware-software co-designed solution solves three key challenges: i) the overlapping sum mapping problem; ii) ineffectual computations and cropped output mapping; and iii) the need for efficient dataflow strategies for resource-constrained edge devices.
We implemented our proposed hardware design on an edge FPGA using the SECDA-TFLite toolkit, and evaluated the performance across a large variety of configurations for TCONV problems, achieving an average speedup of \SynthSpeedUp against a dual-thread ARM CPU.
We also compared MM2IM against other TCONV accelerators for similarly resource-constrained edge FPGAs and achieved at least \atleastGOPsDSPx higher GOPs/DSP compared to the next best accelerator.
Finally, we performed an end-to-end evaluation of the DCGAN and pix2pix models, achieving a \dualModelSpeedUp speedup and \dualModelEnergyreduction energy reduction on average compared with the CPU baseline.
As future work, we plan to further scale down our accelerator design to enable TCONV in devices with lower resources budgets such as micro-controllers.


\section*{Acknowledgments}

This work was partially supported by the UK Engineering and Physical Sciences Research Council (grant EP/R513222/1), the EU Project dAIEDGE (GA Nr 101120726) and the Innovate UK Horizon Europe Guarantee (GA Nr 10090788).




\balance

\bibliographystyle{IEEEtran}
\bibliography{full_bib}

\begin{thebibliography}{10}
\providecommand{\url}[1]{#1}
\csname url@samestyle\endcsname
\providecommand{\newblock}{\relax}
\providecommand{\bibinfo}[2]{#2}
\providecommand{\BIBentrySTDinterwordspacing}{\spaceskip=0pt\relax}
\providecommand{\BIBentryALTinterwordstretchfactor}{4}
\providecommand{\BIBentryALTinterwordspacing}{\spaceskip=\fontdimen2\font plus
\BIBentryALTinterwordstretchfactor\fontdimen3\font minus \fontdimen4\font\relax}
\providecommand{\BIBforeignlanguage}[2]{{%
\expandafter\ifx\csname l@#1\endcsname\relax
\typeout{** WARNING: IEEEtran.bst: No hyphenation pattern has been}%
\typeout{** loaded for the language `#1'. Using the pattern for}%
\typeout{** the default language instead.}%
\else
\language=\csname l@#1\endcsname
\fi
#2}}
\providecommand{\BIBdecl}{\relax}
\BIBdecl

\bibitem{dongAcceleratingSuperResolutionConvolutional2016}
C.~Dong, C.~C. Loy, and X.~Tang, ``Accelerating the {{Super-Resolution Convolutional Neural Network}},'' in \emph{European {{Conference}} on {{Computer Vision} (ECCV)}}, 2016, pp. 391--407.

\bibitem{johnsonPerceptualLossesRealTime2016}
J.~Johnson, A.~Alahi, and L.~Fei-Fei, ``Perceptual {{Losses}} for {{Real-Time Style Transfer}} and {{Super-Resolution}},'' in \emph{European {{Conference}} on {{Computer Vision} (ECCV)}}, 2016, pp. 694--711.

\bibitem{liuGenerativeModelingSmallData2019}
L.~Liu, M.~Muelly, J.~Deng, T.~Pfister, and L.-J. Li, ``Generative {{Modeling}} for {{Small-Data Object Detection}},'' in \emph{2019 {{IEEE}}/{{CVF International Conference}} on {{Computer Vision}} ({{ICCV}})}, 2019, pp. 6072--6080.

\bibitem{gibson_dlas_2024}
P.~Gibson, J.~Cano, E.~J. Crowley, A.~Storkey, and M.~O'Boyle, ``{DLAS: A Conceptual Model for Across-Stack Deep Learning Acceleration},'' in \emph{ACM Transactions on Architecture and Code Optimization}, 2024.

\bibitem{mittalSurveyFPGAbasedAccelerators2020}
\BIBentryALTinterwordspacing
S.~Mittal, ``A survey of {{FPGA-based}} accelerators for convolutional neural networks,'' vol.~32, no.~4, pp. 1109--1139. [Online]. Available: \url{https://doi.org/10.1007/s00521-018-3761-1}
\BIBentrySTDinterwordspacing

\bibitem{zhangDesignMethodologyEfficient2017a}
X.~Zhang, S.~Das, O.~Neopane, and K.~Kreutz-Delgado, ``A {{Design Methodology}} for {{Efficient Implementation}} of {{Deconvolutional Neural Networks}} on an {{FPGA}},'' in \emph{{{arXiv:1705.02583}}}, 2017.

\bibitem{yuUniOPUFPGABasedUniform2020a}
Y.~Yu, T.~Zhao, M.~Wang, K.~Wang, and L.~He, ``Uni-{{OPU}}: {{An FPGA-Based Uniform Accelerator}} for {{Convolutional}} and {{Transposed Convolutional Networks}},'' \emph{IEEE Transactions on Very Large Scale Integration (VLSI) Systems}, pp. 1545--1556, 2020.

\bibitem{changEnergyEfficientFPGABasedDeconvolutional2020}
J.-W. Chang, K.-W. Kang, and S.-J. Kang, ``An {{Energy-Efficient FPGA-Based Deconvolutional Neural Networks Accelerator}} for {{Single Image Super-Resolution}},'' \emph{IEEE Transactions on Circuits and Systems for Video Technology}, pp. 281--295, 2020.

\bibitem{yanGNAReconfigurableEfficient2018a}
J.~Yan, S.~Yin, F.~Tu, L.~Liu, and S.~Wei, ``{{GNA}}: {{Reconfigurable}} and {{Efficient Architecture}} for {{Generative Network Acceleration}},'' \emph{IEEE Transactions on Computer-Aided Design of Integrated Circuits and Systems}, pp. 2519--2529, 2018.

\bibitem{maIntermediateCentricDataflowTransposed2023}
Z.~Ma, T.~Dai, X.~Wei, and G.~Luo, ``An {{Intermediate-Centric Dataflow}} for {{Transposed Convolution Acceleration}} on {{FPGA}},'' \emph{ACM Transactions on Embedded Computing Systems}, 2022.

\bibitem{xuFCNEngineAcceleratingDeconvolutional2018}
D.~Xu, K.~Tu, Y.~Wang, C.~Liu, B.~He, and H.~Li, ``{{FCN-Engine}}: {{Accelerating Deconvolutional Layers}} in {{Classic CNN Processors}},'' in \emph{2018 {{IEEE}}/{{ACM International Conference}} on {{Computer-Aided Design} (ICCAD)}}, 2018, pp. 1--6.

\bibitem{sestitoFPGADesignTransposed2023}
\BIBentryALTinterwordspacing
C.~Sestito, S.~Perri, and R.~Stewart, ``{{FPGA Design}} of {{Transposed Convolutions}} for {{Deep Learning Using High-Level Synthesis}},'' \emph{Journal of Signal Processing Systems}, 2023. [Online]. Available: \url{https://doi.org/10.1007/s11265-023-01883-7}
\BIBentrySTDinterwordspacing

\bibitem{matlabdevsCol2imRearrangeMatrix}
\BIBentryALTinterwordspacing
M.~Devs. Col2im - {{Rearrange}} matrix columns into blocks - {{MATLAB}}. [Online]. Available: \url{https://uk.mathworks.com/help/images/ref/col2im.html}
\BIBentrySTDinterwordspacing

\bibitem{harisSECDATFLiteToolkitEfficient2023}
J.~Haris, P.~Gibson, J.~Cano, N.~Bohm~Agostini, and D.~Kaeli, ``{{SECDA-TFLite}}: {{A}} toolkit for efficient development of {{FPGA-based DNN}} accelerators for edge inference,'' \emph{Journal of Parallel and Distributed Computing}, pp. 140--151, 2023.

\bibitem{radfordUnsupervisedRepresentationLearning2016}
A.~Radford, L.~Metz, and S.~Chintala, ``Unsupervised {{Representation Learning}} with {{Deep Convolutional Generative Adversarial Networks}},'' in \emph{4th {{International Conference}} on {{Learning Representations} (ICLR)}}, 2016.

\bibitem{harisSECDAEfficientHardware2021}
J.~Haris, P.~Gibson, J.~Cano, N.~B. Agostini, and D.~Kaeli, ``{{SECDA}}: {{Efficient Hardware}}/{{Software Co-Design}} of {{FPGA-based DNN Accelerators}} for {{Edge Inference}},'' in \emph{2021 {{IEEE}} 33rd {{International Symposium}} on {{Computer Architecture}} and {{High Performance Computing} (SBAC-PAD)}}, 2021, pp. 33--43.

\bibitem{longFullyConvolutionalNetworks2015}
\BIBentryALTinterwordspacing
J.~Long, E.~Shelhamer, and T.~Darrell. Fully {{Convolutional Networks}} for {{Semantic Segmentation}}. [Online]. Available: \url{http://arxiv.org/abs/1411.4038}
\BIBentrySTDinterwordspacing

\bibitem{liuOptimizingCNNbasedSegmentation2018a}
S.~Liu, H.~Fan, X.~Niu, H.-c. Ng, Y.~Chu, and W.~Luk, ``Optimizing {{CNN-based Segmentation}} with {{Deeply Customized Convolutional}} and {{Deconvolutional Architectures}} on {{FPGA}},'' \emph{ACM Transactions on Reconfigurable Technology and Systems}, pp. 1--22, 2018.

\bibitem{diExploringEfficientAcceleration2020}
X.~Di, H.-G. Yang, Y.~Jia, Z.~Huang, and N.~Mao, ``Exploring {{Efficient Acceleration Architecture}} for {{Winograd-Transformed Transposed Convolution}} of {{GANs}} on {{FPGAs}},'' \emph{Electronics}, p. 286, 2020.

\bibitem{isolaImagetoImageTranslationConditional2018}
P.~Isola, J.-Y. Zhu, T.~Zhou, and A.~A. Efros, ``Image-to-{{Image Translation}} with {{Conditional Adversarial Networks}},'' in \emph{{{arXiv:1611.07004}}}.

\bibitem{changDesignMethodologyEfficient2020}
J.-W. Chang, S.~Ahn, K.-W. Kang, and S.-J. Kang, ``Towards {{Design Methodology}} of {{Efficient Fast Algorithms}} for {{Accelerating Generative Adversarial Networks}} on {{FPGAs}},'' in \emph{2020 25th {{Asia}} and {{South Pacific Design Automation Conference} (ASP-DAC)}}, 2020, pp. 283--288.

\bibitem{sestitoHighLevelSynthesisHardware2022}
C.~Sestito, R.~Stewart, and S.~Perri, ``High-{{Level Synthesis}} of {{Hardware Accelerators}} for {{Deconvolution Engines}},'' pp. 1--4.

\bibitem{wuEfficientFPGABasedDilated2024}
T.-H. Wu, C.~Shu, and T.-T. Liu, ``An {{Efficient FPGA-Based Dilated}} and {{Transposed Convolutional Neural Network Accelerator}},'' vol.~71, no.~11, pp. 5178--5186.

\end{thebibliography}

\balance
\end{document}